\documentclass[aps,pre,twocolumn,superscriptaddress,showpacs]{revtex4}
\usepackage{amsmath}
\usepackage{mathrsfs}
\usepackage{amssymb}
\usepackage{amsfonts}
\usepackage{graphicx}

\begin{document}

\title{Fidelity susceptibility for SU(2)-invariant states}
\author{Xiaoguang Wang}
\email{xgwang@zimp.zju.edu.cn}
\affiliation{Department of Physics and ITP, The Chinese University of Hong Kong, Hong
Kong, China}
\affiliation{Zhejiang Institute of Modern Physics, Department of Physics, Zhejiang
University, HangZhou 310027, P.R. China.}
\author{Shi-Jian Gu}
\affiliation{Department of Physics and ITP, The Chinese University of Hong Kong, Hong
Kong, China}
\date{\today}

\begin{abstract}
We study the fidelity susceptibility of two SU(2)-invariant reduced
density matrices. Due to the commuting property of these matrices,
analytical results for reduced fidelity susceptibility are obtained
and can be applied to study quantum phase transitions in
SU(2)-invariant systems. As an example, we analyze the quantum
criticality of the spin-one bilinear-biquadratic model via the
fidelity approach.
\end{abstract}

\maketitle

\textit{Introduction}---In 1984, Peres introduced the concept of
fidelity to characterize quantum system responses to a
perturbation~\cite{Peres}. It is of fundamental importance when
studying quantum dynamics and has been applied to characterize two
important phenomena in condensed matter theory,
quantum chaos and quantum phase transitions (QPTs) \cite%
{Lloyd,Suncp,Zanardi,fidelity-sspt,HQZhou0701,MFYang07,LGong115114}. It also
becomes an useful concept in quantum information theory \cite{Nielsen1}, and
has been used in study of quantum entanglement theory~\cite{Wei}, quantum
teleportation~\cite{Hofmann}, and transformation of unknown states~\cite{Lars}
etc.

As an indicator of QPTs, various kinds of fidelity has been used in
investigating quantum phase transition point, such as Loschmidt
echo~\cite{Suncp}, ground-state fidelity \cite{GS-fidelity}, the
fidelity of first excited state \cite{FES-fidelity}, operator
fidelity~\cite{operator-fidelity-sspt,op}, and reduced
fidelity~\cite{reduced-fidelity,partial-fidelity} etc. The fidelity susceptibility (FS)~\cite%
{fidelity-sspt}, as the leading term of the fidelity, can be
conveniently used to detect QPTs for its independence of the
concrete values of small perturbations.

Let us briefly introduce quantum fidelity and fidelity susceptibility. For
pure states, fidelity is the absolute value of an overlap of two wave
functions. One important case is the fidelity between the ground state $%
|\psi _{0}(x)\rangle $ of Hamiltonian $H(x)$ and a slightly different one $%
|\psi _{0}(x+\delta )\rangle $,
\begin{equation}
F=|\langle \psi _{0}(x)|\psi _{0}(x+\delta )\rangle |,
\end{equation}%
where $\delta $ is a small deviation. Substituting the expansion
\begin{equation}
|\psi (x+\delta )\rangle =|\psi (x)\rangle +\delta |\dot{\psi}(x)\rangle
+\delta ^{2}/2|\ddot{\psi}(x)\rangle +O(\delta ^{3})
\end{equation}%
into the above equation leads to
\begin{align}
F& =1+\frac{\delta ^{2}}{4}(\langle \psi _{0}|\ddot{\psi}_{0}\rangle
+\langle \ddot{\psi}_{0}|{\psi }_{0}\rangle +2|\langle \psi _{0}|\dot{\psi}%
_{0}\rangle |^{2})  \notag \\
& =1-\frac{\delta ^{2}}{2}(\langle \dot{\psi}_{0}|\dot{\psi}_{0}\rangle
-|\langle \psi _{0}|\dot{\psi}_{0}\rangle |^{2}).
\end{align}%
One may further define the FS as~\cite%
{fidelity-sspt}
\begin{align}
\chi_F & =\lim_{\delta \rightarrow 0}\frac{2(1-F)}{\delta ^{2}}=\langle \dot{%
\psi}_{0}|\dot{\psi}_{0}\rangle -|\langle \psi _{0}|\dot{\psi}_{0}\rangle
|^{2}  \notag \\
& =\sum_{n\neq 0}|\langle \psi _{n}|\dot{\psi}_{0}\rangle |^{2}.
\end{align}%
So, the FS is explicitly written out in terms of eigenstates.
However, for mixed-state case, the corresponding fidelity and FS is
relatively difficult to be achieved. One can use Uhlmann's
fidelity~\cite{Uhlmann}
\begin{equation}
F=\text{tr}\sqrt{\varrho ^{1/2}\widetilde{\varrho }\varrho ^{1/2}}
\label{ffffffffff}
\end{equation}%
for two mixed states $\varrho $ and $\widetilde{\varrho }$ and the
corresponding FS can also be defined as above. In what follows, we
analyze fidelity of SU(2)-invariant mixed states by this definition.

For a many-body quantum state, by tracing out other degree of
freedom but two particles, we have a two-particle reduced-density
matrix which is generally a mixed state. Fidelity between reduced
density matrices is called reduced fidelity~\cite{reduced-fidelity}.
For some interesting physical
models such as the spin-one bilinear-biquadratic model~\cite%
{Affleck,Millet,Xiang,Botet} and spin-half frustrated model, the
reduced-density matrix displays a SU(2) symmetry. In this paper, we
will study the fidelity of SU(2)-invariant state. The symmetry in
this state faciliates greatly our study of fidelity and analytical
results are obtained for the fidelity susceptibility. We also give
an application of the results to study quantum phase transition in
the bilinear-biquadratic model.

\textit{SU(2)-invariant states and FS---} Before proceeding, we make it
clear that if a multi-spin state $\rho $ displays a global SU(2) symmetry ($%
[\rho ,\mathbf{J}]=0$), the two-spin reduced-density matrix also has a SU(2)
symmetry ($[\rho _{12},\mathbf{j}_{1}+\mathbf{j}_{2}]=0$), where $\mathbf{J}=%
\mathbf{j}_{1}+...+\mathbf{j}_{N}$ is the collecting spin operator.
The proof is straightforward. The commutator $[\rho ,\mathbf{J}]=0$
means that
\begin{equation}
\lbrack \rho ,\mathbf{j}_{1}+\mathbf{j}_{2}]=[\mathbf{j}_{3}+...+\mathbf{j}%
_{n},\rho ].
\end{equation}%
After tracing out degree of freedom of spins $3\rightarrow N$, we have
\begin{equation}
\lbrack \rho _{12},\mathbf{j}_{1}+\mathbf{j}_{2}]=\text{Tr}_{3\rightarrow N}[%
\mathbf{j}_{3}+...+\mathbf{j}_{n},\rho ]=0.
\end{equation}

An SU(2)-invariant state of two spins $j_1$ and $j_2$ can be written in the
general form
\begin{equation}  \label{aa}
\rho=\sum_{J=|j_1-j_2|}^{j_1+j_2}\frac{\alpha_J}{2J+1}P_J,
\end{equation}
where $\alpha_J\ge 0$, $\sum_J \alpha_J=1$, and $P_J$ is the projector of
spin-$J$ subspace. Obviously, the density operator has eigenvalues $%
\alpha_J/(2J+1)$ with degeneracy $2J+1$.

One key observation from the above equation is that two different
SU(2)-invairant density matrices $\rho $ and $\tilde{\rho}$ commute with
each other. Thus, they can be diagonalized simutaneously, and the fidelity
between them are given by.
\begin{equation}
F=\sum_{k=1}^{j_{1}j_{2}}\sqrt{\lambda _{k}\widetilde{\lambda }_{k}},
\label{ff}
\end{equation}%
where $\lambda _{k}$'s and $\widetilde{\lambda }_{k}$'s are the
eigenvalues of $\rho $ and $\widetilde{\varrho }$, respectively.
Since zero eigenvalues have no contribution to $F$, we only need to
consider the nonzero ones. In the following, the subscript $k$ in
$\sum_{k}$ only denotes nonzero eigenvalues of $\rho $.

Now we calculate fidelity of two slightly different density matrices
$\rho (x)$ and $\rho (x+\delta )$ as a function of parameter $x$,
where $\delta $ is a small change of $x$. It is noticed that, for a
small change $\delta ,$ we have
\begin{equation}
{\lambda _{k}(x+\delta )}\simeq \lambda _{k}+\left( \partial _{x}\lambda
_{k}\right) \delta +\left( \partial _{x}^{2}\lambda _{k}\right) \delta ^{2}/{%
2}+O\left( \delta ^{3}\right) .
\end{equation}%
Substituting this expression into Eq.~(\ref{ff}) leads to the fidelity given
by
\begin{equation}
F=1-\frac{\delta ^{2}}{2}\sum_{i}\frac{\left( \partial _{x}\lambda
_{k}\right) ^{2}}{4\lambda _{k}}.
\end{equation}%
In deriving the above equation, we have used $\sum_{i}\lambda
_{i}\equiv {1}$ and $\sum_{i}\partial _{\alpha }\lambda
_{i}=\sum_{i}\partial _{\alpha }^{2}\lambda _{i}=0.$ Therefore,
according to the relation between fidelity and FS $F=1-\chi {\delta
^{2}/2}$~\cite{fidelity-sspt}, the
FS $\chi_{F}$ corresponding to the matrix $%
\rho $ is obtained as
\begin{equation}
\chi_F=\sum_{k}\frac{\left( \partial _{x}\lambda _{k}\right)
^{2}}{4\lambda _{k}}.  \label{fs}
\end{equation}%
This expression of fidelity susceptibility is valid for any commuting
density matrices. It depends on nonzero eigenvalues of $\rho $ and their
first-order derivatives.

Applying the Eq.~(\ref{fs}) to the SU(2)-invariant state $\rho $ (\ref{aa}),
one obtains the FS for $\rho $ as
\begin{equation}
\chi_F =\sum_{J=j_{2}-j_{1}}^{j_{1}+j_{2}}\frac{(\partial _{x}\alpha _{J})^{2}%
}{4\alpha _{J}},  \label{chi}
\end{equation}%
where we assumed $j_{2}>j_{1}$ without loss of generality. Now, we
consider the following case of $j_{1}=1/2$ and $j_{2}\geq 1/2$. As
$\alpha _{j_{2}-1/2}+\alpha _{j_{2}+1/2}=1$, Eq.~(\ref{chi}) reduces
to
\begin{equation}
\chi_F =\frac{(\partial _{x}\alpha _{j_{2}-1/2})^{2}}{4\alpha
_{j_{2}-1/2}(1-\alpha _{j_{2}-1/2})}.  \label{chichi}
\end{equation}%
Parameter $\alpha _{j_{2}-1/2}$ can be written in terms of expectation of
Heisenberg interaction on $\rho $, $\langle \mathbf{j}_{1}\cdot \mathbf{j}%
_{2}\rangle $, i.e., ~\cite{Schliemann}
\begin{equation}
\alpha _{j_{2}-1/2}=\frac{1}{2j_{2}+1}(j_{2}-2\langle \mathbf{j}_{1}\cdot
\mathbf{j}_{2}\rangle ).
\end{equation}%
Thus, Eq.~(\ref{chichi}) can be reexpressed in the following form
\begin{equation}
\chi_F =\frac{(\partial _{x}\langle \mathbf{j}_{1}\cdot
\mathbf{j}_{2}\rangle )^{2}}{(j_{2}-2\langle \mathbf{j}_{1}\cdot
\mathbf{j}_{2}\rangle )(j_{2}+1+2\langle \mathbf{j}_{1}\cdot
\mathbf{j}_{2}\rangle )} \label{chichichi}
\end{equation}%
We see that for the SU(2)-invariant state, the FS is completely determined
by the expectation value of Heisenberg interaction and its first-order
derivative. If we consider the case of two qubits, then the above equation
reduces to
\begin{equation}
\chi_F =\frac{4(\partial _{x}\langle \mathbf{j}_{1}\cdot
\mathbf{j}_{2}\rangle )^{2}}{(1-4\langle \mathbf{j}_{1}\cdot
\mathbf{j}_{2}\rangle )(3+4\langle \mathbf{j}_{1}\cdot
\mathbf{j}_{2}\rangle )},  \label{chii}
\end{equation}%
which is just the FS obtained in Ref.~\cite{Xiong} via a different approach.

Now, we study the case of two qutrits, i.e., two spin ones. From
Eq.~(\ref{aa}), one has
\begin{align}
\alpha _{0}& =\langle P_{0}\rangle =\langle P_{12}\rangle ,  \notag
\label{e1} \\
\alpha _{1}& =\langle P_{1}\rangle =\frac{1}{2}(1-\langle S_{12}\rangle ),
\notag \\
\alpha _{2}& =\langle P_{2}\rangle =\frac{1}{2}(1-2\langle P_{12}\rangle
+\langle S_{12}\rangle ),
\end{align}%
where
\begin{align}
P_{12}& =\frac{1}{3}[(\mathbf{j}_{1}\cdot \mathbf{j}_{2})^{2}-1],  \notag
\label{e2} \\
S_{12}& =\mathbf{j}_{1}\cdot \mathbf{j}_{2}+(\mathbf{j}_{1}\cdot \mathbf{j}%
_{2})^{2}-1,
\end{align}%
are singlet projection operator and swap operator, respectively.
Substituting Eqs.~(\ref{e1}) and (\ref{e2}) into Eq.~(\ref{chi})
leads to the FS for two qutrits
\begin{align}
\chi_F & =\frac{1}{4}\left[ {\Huge \frac{(\partial _{x}\langle
P_{12}\rangle )^{2}}{\langle P_{12}\rangle }+\frac{(\partial
_{x}\langle S_{12}\rangle
)^{2}}{2(1-\langle S_{12}\rangle )}}\right.   \notag  \label{e3333} \\
& \left. +\frac{(\partial _{x}\langle S_{12}\rangle -2\partial _{x}\langle
P_{12}\rangle )^{2}}{2(1-2\langle P_{12}\rangle +\langle S_{12}\rangle )}%
\right] {\huge ,}
\end{align}%
The FS is detrmined by two expectation values $\langle P_{12}\rangle $ and $%
\langle S_{12}\rangle $ and their first-order derivatives. Below, we will
apply this formula to the study of the bilinear-biquadratic model.

\textit{Applications to spin-one systems}--- Spin Heisenberg chains attract
more attention since Haldane predicted that the one-dimensional chain has a
spin gap for integer spins~\cite{Haldane}. In these studies, the
bilinear-biquadratic model has played an important role~\cite%
{Affleck,Millet,Xiang,Botet}. The corresponding Hamiltonian is given by
\begin{eqnarray}
H_{BB} &=&\sum_{i=1}^{N}\cos \theta \left( \mathbf{j}_{i}\cdot \mathbf{j}%
_{i+1}\right) +\sin \theta \left( \mathbf{j}_{i}\cdot \mathbf{j}%
_{i+1}\right) ^{2}, \\
&=&\sum_{i=1}^{N}[\cos \theta S_{i,i+1}+3(\sin \theta -\cos \theta
)P_{i,i+1}]+N\sin \theta .  \notag
\end{eqnarray}%
In deriving the last equality, we have used Eq.~(\ref{e2}). Here, $\mathbf{j}%
_{i}$ denotes spin-1 operator at site $i$, and we have assumed the
periodic boundary conditions. The Hamiltonian exhibits a SU(2)
symmetry, and displays very rich quantum phase
diagrams~\cite{Schollwock}.

From Hellmann-Feymann theorem for ground state, one can easily find that
\begin{align}
\langle P_{12}\rangle & =\frac{1}{3}(\sin \theta e_{0}+\cos \theta
e_{0}^{\prime }-1),  \notag \\
\langle S_{12}\rangle & =(\cos \theta +\sin \theta )e_{0}+(\cos \theta -\sin
\theta )e_{0}^{\prime }-1,
\end{align}%
and the their first-order derivatives
\begin{align}
\langle P_{12}\rangle ^{\prime }& =\frac{\cos \theta }{3}(e_{0}+e_{0}^{%
\prime \prime }),  \notag \\
\langle S_{12}\rangle ^{\prime }& =({\cos \theta -\sin \theta }%
)(e_{0}+e_{0}^{\prime \prime }),  \notag \\
\langle S_{12}\rangle ^{\prime }-2\langle P_{12}\rangle ^{\prime }& =\left(
\frac{\cos \theta }{3}-\sin \theta \right) (e_{0}+e_{0}^{\prime \prime }).
\end{align}%
Here, $e_{0}$ denotes the ground-state energy per site. Substituting the
above two equations into (\ref{e3333}), one obtains the FS in terms of $e_{0}
$, $e_{0}^{\prime }$, and $e_{0}^{\prime \prime }$ as follows
\begin{align}
\chi _{F}& =\frac{(e_{0}+e_{0}^{\prime \prime })^{2}}{4}\left[ \frac{\cos
^{2}\theta }{3(\sin \theta e_{0}+\cos \theta e_{0}^{\prime }-1)}\right.
\notag  \label{la} \\
& +\frac{(\cos \theta -\sin \theta )^{2}}{2[2-(\cos \theta +\sin \theta
)e_{0}-(\cos \theta -\sin \theta )e_{0}^{\prime }]}  \notag \\
& \left. +\frac{(\cos \theta -3\sin \theta )^{2}}{6[2+(3\cos \theta +\sin
\theta )e_{0}+(\cos \theta -3\sin \theta )e_{0}^{\prime }]}\right]
\end{align}

One key observation is that the numerators of the above two expressions
happen to be proportional to $(e_{0}+e_{0}^{\prime \prime })^{2}$. Then, we
infer that if the second derivative of the ground-state energy is singular
at the critical point, the FS is singular too. On the other hand, it is
known that the divergence of the second derivative of the ground-state
energy reflects the second-order QPTs of the system, which is shown in Ref.
\cite{GS-fidelity} explicitly as
\begin{equation*}
\partial _{\alpha }^{2}e_{0}=\sum_{n\neq 0}^{N}\frac{2\left\vert \langle
\Psi _{n}|\partial _{\alpha }H|\Psi _{n}\rangle \right\vert ^{2}}{%
N(E_{0}-E_{n})},
\end{equation*}%
where $|\Psi _{n}\rangle $ is the eigenvector corresponding to the
eigenvalue $E_{n}$. It shows that the vanishing energy gap in the
thermodynamic limit can lead to the singularity of the the second
derivative of the ground-state energy. Therefore, the two-spin FSs
can exactly reflects the second-order QPTs of the global system in
this model.

\begin{figure}[ptb]
\begin{center}
\includegraphics[width=8cm] {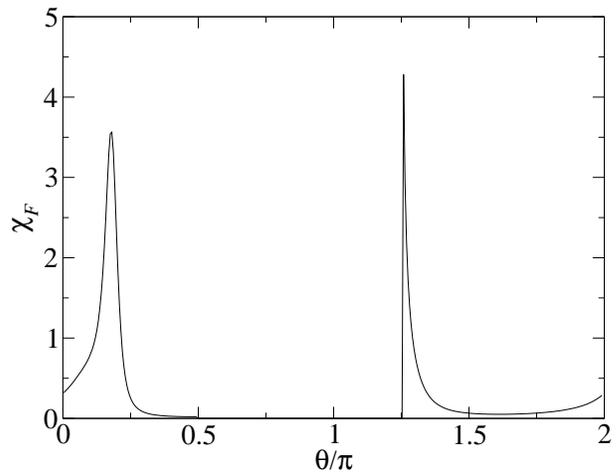}
\end{center}
\caption{Ground-state fidelity susceptibility as a function of $\protect%
\theta/\protect\pi$ in the bilinear-biquadratic model.}
\label{fig-fds-limit}
\end{figure}

We use the exact-diagonalization method to calcuate the ground-state energy
and then numerical results of FS is obtained from Eq.(\ref{la}) . In Fig. 1,
we plot the FS as a function of $\theta $ for a sample of 12 spins. We
observe a sharp decrease of the FS around $\theta =\pi /4$, which separate
the Haldane phase ($-\pi /4<\theta <\pi /4$) and the trimerized phase ($\pi
/4<\theta <\pi /2$). This may imply that a QPT occurs. For $\pi /2<\theta
<5\pi /4$, the ground state is ferromagnetic and degenerate. In this range,
the FS is zero. However, at $\theta =7\pi /4$, corresponding to a QPT point
separating dimerized phase ($5\pi /4<\theta <7\pi /4$) and Haldane phase,
one cannot find any anomalous behaviors of the FS.

\textit{Conclusions}---We have studied the FS in SU(2)-invariant
states and have obtained exact analytical expression of the FS for
any spins $j_{1}$ and $j_{2}$. This implies that the results are
applicable to not only equal-spin but also mixed-spin systems.
Furthermore, one can use the FS to study properties of
SU(2)-invariant physical systems in a finite-temperature thermal
state. As an application, we have studied relations between the FS
and QPTs in the bilinear-biquadratic model. For this model, one can
infer that the two-spin FS can exactly reflects the second-order
QPTs of the global system. Here, we restrict us to study
SU(2)-invariant states of two spins, after tracing out other spins
of a many-body states. One challenge for further investigation is to
study $N$-spin ($N\geq 3$) SU(2)-invariant reduced density matrix.

\textit{Acknowledgements} This work was supported by the Program for New
Century Excellent Talents in University (NCET), the NSFC with grant No.
90503003, the State Key Program for Basic Research of China with grant No.
2006CB921206, the Specialized Research Fund for the Doctoral Program of
Higher Education with grant No. 20050335087, and the Direct grant of CUHK
(A/C 2060344). X. Wang acknowledge the support from C. N. Yang fellowship
via CUHK.

\end{document}